\documentclass[aps,twocolumn,groupedaddress,showpacs,amssymb,prl,floatfix,preprintnumbers]{revtex4}
\usepackage{graphicx,color}

\newcommand{\slrr}      {$T_1^{-1}$}

\newcommand{\si}        {$^{29}$Si}

\newcommand{\cecoin}    {CeCoIn$_5$}

\newcommand{\tn}     {$T_{\rm N}$}

\newcommand{\slrrtext}  {spin lattice relaxation rate}
\newcommand{\urusi}     {URu$_2$Si$_2$}

\newcommand{\urusirh}     {U(Ru$_{1-x}$Rh$_x$)$_2$Si$_2$}

\begin{document}

\thispagestyle{myheadings}

\title{Uncovering the Hidden Order in \urusi\ by Impurity Doping}

\author
{S.-H. Baek$^1$, M. J. Graf$^1$, A. V. Balatsky$^1$, E.
D. Bauer$^1$,
J. C. Cooley$^1$, J. L. Smith$^1$, N. J. Curro$^{2}$ \email{curro@physics.ucdavis.edu}}

 \affiliation{$^{1}$Los Alamos National Laboratory, Los Alamos, NM 87545, USA\\
 $^{2}$Department of Physics, University of California, Davis, CA 95616, USA}


\date{\today}

\begin{abstract}

We report the use of impurities to probe the hidden order parameter of the strongly correlated metal \urusi\ below the
transition temperature $T_0\sim$ 17.5 K.  The nature of this order parameter has eluded researchers for more than two decades, but is accompanied by
the development of a partial gap in the single particle density of states
that can be detected through measurements of the electronic specific heat and nuclear spin-lattice relaxation rate.
We find that impurities in the hidden order phase give rise to local patches of
antiferromagnetism. An analysis of the coupling between the antiferromagnetism and the hidden order reveals that the former is not a competing order parameter but rather a parasitic effect of the latter.
\end{abstract}


\maketitle

The heavy fermion \urusi\ has received considerable attention because it undergoes a phase transition to a state which is poorly understood. The strong interactions between the U 5f electrons and the delocalized conduction electrons give rise to an enhanced Sommerfeld coefficient $\gamma = 180$ mJ/mol K$^{-2}$ and two phase transitions at low temperature:
the hidden order (HO) transition at $T_0\sim 17.5$ K gaps approximately 70\% of the Fermi surface area
and a superconducting transition at $T_c \sim 1.4 $ K emerges from the remaining charge carriers \cite{PalstraURSdiscovery,KoharaURSinhomogeneity}.
The large entropy associated with the HO phase transition is suggestive of spin density wave order, yet direct spin probes have shown no evidence of intrinsic magnetic order in pure crystals.
Although the HO phase of \urusi\ is not itself magnetic, this phase is closely related to antiferromagnetism (AF) of the U electron spins.  Early neutron scattering and muon spin rotation ($\mu$SR) studies reported a tiny ordered magnetic moment of 0.03 $\mu_B$/U in pure \urusi, which led to the concept of small moment antiferromagnetism (SMAF) \cite{BroholmURS,MacLaughlinMSRURS}.  However, later $\mu$SR and nuclear magnetic resonance (NMR) measurements tell a quite different story \cite{LukeURSmuSR,KoharaURSinhomogeneity}.  They
reveal an inhomogeneous coexistence between small regions of antiferromagnetic order and hidden order in pure \urusi, with a relative fraction that tends toward bulk AF under pressure \cite{BroholmURS,KoharaURSinhomogeneity,LukeURSmuSR}.
Substituting Rh for Ru in \urusi\ leads to a suppression of the long range hidden order, and recent neutron scattering studies revealed large moment AF coexisting with the hidden order for large Rh concentrations \cite{URSRhneutrons}.

\begin{figure}
\includegraphics[width=\linewidth]{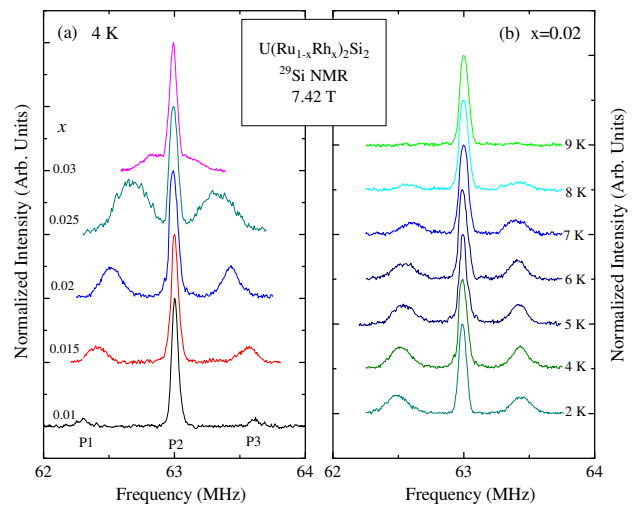}
\caption{\label{fig:spectra} (color online) \si\ spectra in \urusirh\ as a function of
temperature and Rh concentration. The spectra were obtained by summing the
Fourier transforms of Hahn echoes for several different frequencies in a fixed
external field along the $c$-direction.
(a) Spectra as a function of Rh concentration $x$ at 4 K. Spectra were
normalized to the height of central line. The spectra are consistent with commensurate antiferromagnetic ordering with moments aligned (anti)parallel to the $c$-direction.
(b) Temperature dependence of the spectra for fixed $x=0.02$. The antiferromagnetic signature disappears gradually near 9 K without change of the ordered moment.}
\vspace{-0.15in}
\end{figure}

\begin{figure}
\includegraphics[width=0.9\linewidth]{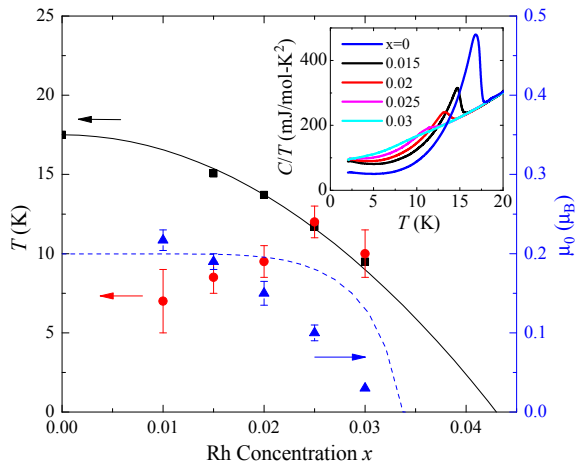}
\caption{\label{fig:moment} (color online) The coexistence phase diagram of the HO (black square; left axis) and
the AF order (red circle; left axis) as a function of doping.
Blue triangles (right axis) indicate the magnetic moment of the AF order.
The transition of the HO order is determined by the specific heat, while the induced AF transition is determined by the disappearance
of the AF peaks.
The lines are the calculated $T_0(x)$ (solid black) and $\mu_0(x)$ (dashed blue) using the GL model
described in the text. The calculated curves are renormalized by the critical concentration $x_c$ \cite{GL_parameters}.
INSET: Specific heat over temperature, $C/T$, as a function of temperature $T$ and doping $x$. There is no sign of a
bulk phase transition at $T_{\rm AF}(x)$.}
\end{figure}

In order to investigate the microscopic effects of the Rh dopants on the hidden order phase and to characterize the emergent AF in \urusirh\ we have measured the \si\ NMR spectrum as a
function of temperature and Rh concentration.  Fig. \ref{fig:spectra} shows a
series of such spectra.  The resonance frequency of the \si\ (nuclear spin $I=\frac{1}{2}$) is given by $f=\gamma H_0(1+K)$, where $\gamma$ is the gyromagnetic ratio of the \si, $H_0$ is the applied external field (7 T), and $K$ is the Knight shift arising from the hyperfine coupling between the nuclear and electron spins in the solid.  Aside from a slight suppression of $K$ connected with the opening of the gap,
there is no visible change of the spectrum at the hidden order transition, $T_0$.
However at a lower temperature \tn\ we find the emergence of two satellite peaks on either side of the central resonance.  These satellites arise because of the
presence of a static internal hyperfine field, $H_{\rm hf}$, associated with commensurate
antiferromagnetic order with moments pointing along (001). The nuclei resonate in the local field $\mathbf{H}_0+\mathbf{H}_{\rm hf}$, where $H_{\rm hf}=A\mu_0$, $\mu_0$ is the ordered U spin moment and  $A$ is the hyperfine coupling. We find that $A$ is unchanged from the pure compound (3.6 kOe/$\mu_B$) \cite{kohoriURu2Si2},
and can therefore directly measure the antiferromagnetic order parameter, $M(T,x)\sim \mu_0(T,x)$, shown in Fig. \ref{fig:moment}.

\begin{figure}
\includegraphics[width=0.8\linewidth]{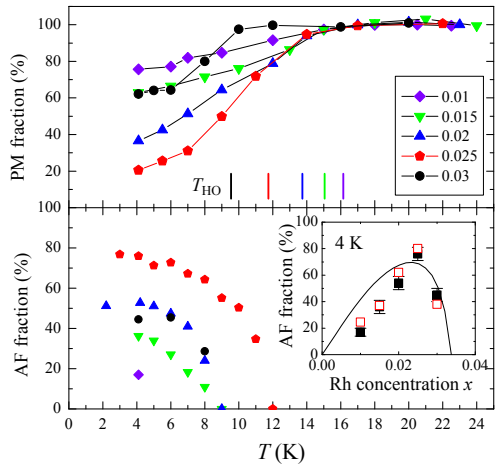}
\caption{\label{fig:fraction} (color online) The spectral weight of the AF signal as a function of temperature and doping.
Upper panel: The paramagnetic (PM) fraction, defined as the relative intensity (area)
of the central line. Spectral intensities were corrected for the Boltzmann factor, and normalized to the high temperature values.
Lower panel: The fractional area under the AF satellites relative to the area of the central
PM peak in the spectra (Fig. \ref{fig:spectra}).
INSET: The nonmonotonic behavior of the AF fraction
versus doping at 4 K. Filled squares are the measured fraction as shown in
lower panel, while empty squares are the indirect results of the lost fraction of the PM peak in the upper panel.
The solid line is the calculated average magnetization (scaled AF fraction)
discussed in the text.}
\end{figure}

    The spectra in Fig. \ref{fig:spectra} reveal an \textit{inhomogeneous} mixture of
antiferromagnetic (satellite peaks) and hidden order (central peak) regions below $T_N$. We see no specific heat anomaly or critical slowing down at $T_N$, suggesting that this transition is not a new thermodynamic phase, but rather a crossover to an inhomogeneous coexistence \cite{flouquetURSpressure}. The volume fraction of antiferromagnetic
domains, shown in Fig. \ref{fig:fraction}, varies with both temperature and
doping. The antiferromagnetic fraction saturates at low temperature at a maximum of 80\% at  $x = 0.025$.  These observations suggest that the antiferromagnetic domains nucleate
around the Rh dopants, forming patches with a radius $\xi_{\rm AF}$ on the order of two to three lattice spacings at zero temperature. The satellites in the spectrum
arise from nuclei within these patches of AF, whereas the central
resonance arises from nuclei outside. The percolation limit is reached at
$x=x_c$ where the antiferromagnetic patches overlap.

\textit{A priori} these results imply that the AF,
characterized by the order parameter ${\bf M}$, competes with the hidden order,
characterized by an order parameter $\Psi$. Much like in the vortex cores of
the cuprates, a competing antiferromagnetic order parameter can emerge in
spatial regions where the dominant superconducting order parameter is
suppressed locally \cite{demler,fradkinhalos}. Microscopically the impurities can create local strains that may stabilize $M$ in the vicinity of the Rh \cite{mydoshstrains}. However, if this were the case, then $\mu_0(0,x)$
should increase with doping and long-range antiferromagnetic order would develop above the
percolation threshold at $x_c$ \cite{demlerSCandSDW}. For example, a competing order parameter is stabilized in Cd doped
\cecoin\ where antiferromagnetic droplets are nucleated at Cd dopants and
 long-range order develops when these droplets overlap \cite{CeCoIn5CdDroplets}.  However, in \urusirh\ detailed measurements of the specific heat  as a function of both temperature and doping (see inset of Fig. \ref{fig:moment}) show no
evidence of a second phase transition associated with long-range
antiferromagnetic order, either within the hidden order phase or outside the
phase when $T_0=0$. Furthermore, as seen in Figs. \ref{fig:spectra} and \ref{fig:moment} the antiferromagnetic order parameter, ${\bf M}(0,x)\sim\mu_0(x)$, vanishes before $\Psi$ does. In
fact, we find that ${\bf M}(0,x)$ scales with $T_0(x)$ (Fig. \ref{fig:moment}), suggesting that the
AF is \textit{controlled} by the hidden order and never exists on its own as true long-range
order but rather as a parasitic effect within the hidden order phase. It is possible that there are in fact two competing effects with Rh doping: local strains that stabilize $M$ and modifications to the electronic structure from the excess carriers introduced by Rh that destabilize both $M$ and $\Psi$ simultaneously.  In this case there is no reason for $M(x)$ and $\Psi(0,x)$ to have the same behavior, and the simultaneous disappearance of both order parameters implies an unlikely coincidence.

Recently Elgazzar~\cite{Elgazzar2008} has suggested that the HO is a dynamic phenomenon in which the
Fermi surface is partially gapped to a commensurate antiferromagnetic state that becomes static under pressure.
The Rh dopants might then serve to pin the local fluctuations of $M_z$, giving rise to local static patches.  Once again, it is not clear why $M(x)$ should track $T_0(x)$ and the HO is completely suppressed when the local patches overlap.  It is possible that the Rh doping simultaneously pins the fluctuations and destabilizes the HO via modifications of the electronic structure. As argued above, though, this scenario requires an unlikely coincidence.  It also is unclear why the pinning would take place only within a few lattice spacings of the dopant even though the HO is a long range phenomenon.  Furthermore, we note that there is little difference in the temperature dependence of the \slrrtext\ measured in the regions of the bulk (central peak) versus the AF droplets (satellite peaks), suggesting the absence of either a dynamic phenomenon or competing order parameter.

\begin{figure}
\includegraphics[width=\linewidth]{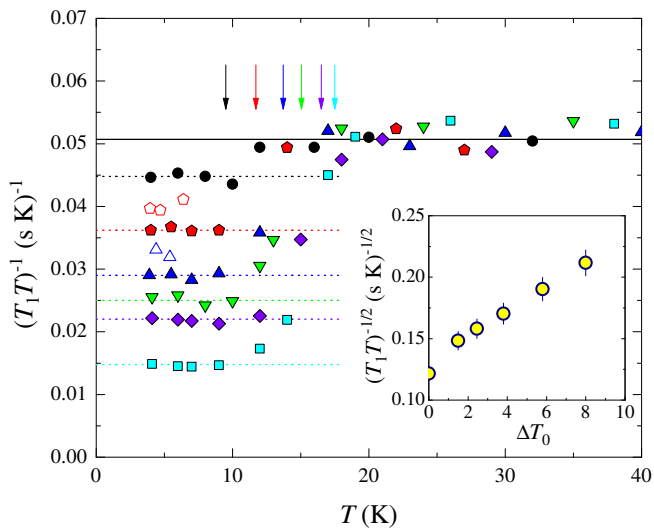}
\caption{\label{fig:T1}
(color online) $(T_1T)^{-1}$ as a function of temperature and doping concentration $x$. Data are shown for
$x=0$ (solid square, cyan),
$x=0.01$ (solid diamond, purple),
$x=0.015$ (solid down triangle, green),
$x=0.02$ (solid up triangle, blue),
$x=0.025$ (solid hexagon, red), and
$x=0.03$ (solid circle, black); open symbols correspond to the antiferromagnetic satellites  at
$x=0.02$ (open up triangle, blue),
and $x=0.025$ (open hexagon, red).  The solid and dotted lines are guides to the eye, and the colored arrows indicate $T_0(x)$. INSET:  $(T_1T)^{-1/2}$ at $T=4$ K versus $\Delta T_0$, revealing the increase in $N(0)$ as the gap is filled by impurity states.} 
\vspace{-0.15in}
\end{figure}

In fact, the observed correlation between the AF ordered moment and the HO gap suggest that the antiferromagnetic patches are an epiphenomenon that is a direct consequence of the local suppression
of the hidden order in the vicinity of the dopants.  We propose that
the antiferromagnetic order is coupled to the spatial derivatives of
$\Psi(\mathbf{r})$.  To interpret the results we used the Ginzburg-Landau (GL) \cite{SigristSCREview}
free energy functional of the combined system that can be written as
$F[\Psi, {\bf M}] = F_{HO} + F_{AF} + F_C$, with
$ F_{HO}[\Psi] =
 a_1(T-T_0)\Psi^2 + \frac{1}{2} b_1\Psi^4 + \kappa_1 |\nabla\Psi|^2  + V \delta({\bf r}) \Psi^2$,
$ F_{AF}[{\bf M}] =
 a_2|{\bf M}|^2+\frac{1}{2} b_2|{\bf M}|^4 + \kappa_2 \left( (\nabla M_x)^2 + (\nabla M_y)^2 + (\nabla M_z)^2 \right)$,
with GL coefficients
$a_1, a_2,  b_1, b_2 >0$,
and impurity potential $V$.
The coupling term is
$F_{C}[\Psi,{\bf M}] =
 g_1\Psi^2 |{\bf M}|^2 + g_2 |{\bf M}|^2 |\nabla\Psi|^2  + g_3 |{\bf M} \cdot \nabla\Psi|^2$.
  The consequences of the first coupling term $g_1$ have been discussed before \cite{shah2000}, while terms $g_2$ and $g_3$ give rise to nucleation of inhomogeneous antiferromagnetic order
around the impurity site where the hidden order is suppressed. Since there is no
experimental evidence for long-range antiferromagnetic order in the undoped system at zero pressure,
($a_2, g_1 \geq 0$ as well as $ b_1, b_2 >0$),
the only way to stabilize a local solution of ${\bf M}({\bf r})$ around an impurity is
by demanding that $g_2,g_3<0$.
To simplify our discussion, we consider only the coupling term
$g_3$ and choose ${\bf M} = (0,0,M)$ along the applied magnetic field.
The effect produced by a $g_2$ term would be similar to that of $g_3$ and will be neglected.
From our analysis it follows that if the hidden order is locally suppressed at the Rh dopants, then
AF naturally emerges in regions near Rh atoms.  The length scale for the recovery
of the hidden order, the coherence length $\xi(T)$, will then determine the
spatial extent of the antiferromagnetic patches and the percolation threshold
then corresponds to a suppression of the long-range hidden order.
Assuming for simplicity that the suppression occurs periodically,
then the suppression of the hidden order transition temperature
will be
$\Delta T_0 = T_0^0 - T_0 \sim q^2 \kappa_1/a_1$,
where $T_0^0$ is the transition temperature of
the undoped system and the  wave vector ${\bf q}$ describes
the spatial modulation of the hidden order parameter.
Thus, to leading order the $T_0$ suppression
will be proportional to the square of the Rh concentration, as experimentally observed,
instead of the usual linear dependence for impurity-averaged theories.
Furthermore, as the hidden order $\Psi$ is gradually suppressed by Rh dopants and
$T_0$ diminishes, the induced (or parasitic) antiferromagnetic order $M$ will decrease as well.
If we treat $M$ as as small perturbation to $\Psi$, then the maximum value
$M_0$ at the impurity site will decrease according to
$M_0(T)^2  \sim  (2^{-1} |g_{3}|  \xi(T)^{-2}\Psi_0(T)^2 - 2{a_2})/{b_2}$,
where the uniform solution of the unperturbed hidden order is $\Psi_0(T)^2 = -a_1(T)/b_1$.  This trend is clearly visible in the data in Fig.~2.

In order to characterize the low energy density of states associated with these localized states
near the Rh impurities, we have measured the  nuclear \slrrtext, \slrr,  as a function of temperature and doping both within and outside of the antiferromagnetic patches. As seen in  Fig. \ref{fig:T1}, $(T_1T)^{-1}\sim N(0)^2$ is suppressed below $T_0$ because of the development of the partial gap in the density of states (DOS) $N(0)$ at the Fermi surface.  With increasing doping, $(T_1T)^{-1}$ increases monotonically within the hidden order phase. This behavior is very similar to the effect of impurities in unconventional superconductors, suggesting that the Rh impurities induce extra states at low energies \cite{balatskyRMP,ouazi}. In this case, we expect $(T_1T)^{-1}(x)\sim N^2(0,x)\sim (\Delta T_0)^2(x)$, consistent with our observations (INSET, Fig. \ref{fig:T1}).
Indeed, $(T_1T)^{-1}$ is faster at the antiferromagnetic satellites in the spectrum, suggesting an excess local DOS within the droplets.
In the case of \urusi\ there are multiple bands and one expects generally two distinct scenarios for the gap to fill up. The first corresponds to a gap in the low energy states for all bands, in which case the impurity doping would fill up the DOS for all the bands. A second possibility is that some of the states remain gapless below $T_0$ while others develop a full gap, which is consistent with the specific heat anomaly.   In this case the impurity induces intragap states in the gap and essentially does not affect the DOS of the ungapped states. We expect the latter to be realized here. This behavior is also  consistent with a subsequent superconducting transition observed at lower temperatures. To test this scenario one would need to observe the DOS in \urusi\ as a function of  Rh in tunneling experiments, like scanning tunneling spectroscopy.

The fact that the AF is manifest only through the spatial gradients of the hidden order rules out theories of
orbital AF and helicity order \cite{chandraURS,URShelicity}. Rather, the hidden order appears to involve compensated spin polarizations on multiple sites such as the triple-spin correlator scenario or an unconventional multi-band spin density wave \cite{URStriplespin,Balatsky2009}. Similar effects are well known in the study of the NMR hyperfine field at the oxygen sites in the doped high temperature superconducting cuprates: Zn or Ni impurities substituted at the Cu sites locally perturb the staggered antiferromagnetic order of the Cu 3d spins, giving rise to finite hyperfine fields at the O sites \cite{alloulreview}. In the absence of impurities, the hyperfine field at the O site vanishes by symmetry. In \urusirh, gradients of the HO parameter may lead to non-cancelation of the net spin per U site, giving rise to the static $M_z$ that we observe.  Our results are consistent with induced
magnetism $M$ being {\em commensurate} with the lattice, while the hidden order $\Psi$ is very likely {\em incommensurate}, as was argued by Wiebe et al. We point out that this discussion implies that Rh doping induces the conversion of  HO to commensurate AF state within each droplet.
If indeed the HO state represents an incommensurate charge density wave (CDW), as argued in \cite{Balatsky2009} then one would expect that impurities induce spin dependent scattering that converts CDW order into magnetic excitations and in addition modifies the momentum of the density wave to make it commensurate.
Thus it remains a fascinating theoretical and experimental puzzle to explain the sudden conversion of incommensurate hidden order into commensurate antiferromagnetic order in
the presence of disorder and possibly pressure.


\begin{thebibliography}{23}
\bibitem{GL_parameters}
In our calculations, we used GL parameters
$a_1=b_1=b_2=\kappa_1=1$, and  $a_2=0.02, \kappa_2=0.05$,
with coupling coefficients $g_1=g_2=0$ and $g_3= -0.5$.
\end{thebibliography}

\end{document}